\documentclass[showpacs,aps,graphicx,twocolumn]{revtex4}

\usepackage{graphicx,CJK}

\usepackage{dcolumn}
\usepackage{bm}

\usepackage{color}

\usepackage[utf8]{inputenc}
\usepackage[T1]{fontenc}
\usepackage{mathptmx}
\usepackage{picinpar}

\begin{document}

\begin{CJK*}{GBK}{song}

\title{Coherent and incoherent theories for photosynthetic energy transfer} 

\author{Ming-Jie Tao,$^{a,b,*}$ Na-Na Zhang,$^{b,}$\footnote{These authors contributed equally to this work.} Peng-Yu Wen,$^{b}$ Fu-Guo Deng,$^{b,c}$ Qing Ai,$^{b,}$\footnote{aiqing@bnu.edu.cn} Gui-Lu Long$^{a,}$\footnote{gllong@tsinghua.edu.cn}}

\address{$^{a}$Department of Physics, Tsinghua University, Beijing 100084, China    \\
$^{b}$Department of Physics, Applied Optics Beijing Area Major Laboratory, Beijing Normal University, Beijing 100875, China\\
$^{c}$NAAM-Research group, Department of Mathematics, Faculty of Science, King Abdulaziz University, Jeddah 21589, Saudi Arabia}

\date{\today}

\begin{abstract}
There is a remarkable characteristic of photosynthesis in nature, that is, the energy transfer efficiency is close to $100\%$. Recently, due to the rapid progress made in the experimental techniques, quantum coherent effects have been experimentally demonstrated.
Traditionally, the incoherent theories are capable of calculating the energy transfer efficiency, e.g., (generalized) F\"{o}rster theory and modified Redfield theory (MRT). However, in order to describe the quantum coherent effects in photosynthesis, one has to exploit coherent theories,
such as hierarchical equation of motion (HEOM), quantum path integral, coherent modified Redfield theory (CMRT), small-polaron quantum master equation, and general Bloch-Redfield theory in addition to the Redfield theory. Here, we summarize the main points of the above approaches, which might be beneficial to the quantum simulation of quantum dynamics of exciton energy transfer (EET) in natural photosynthesis, and shed light on the design of artificial light-harvesting devices.
\end{abstract}

\maketitle  


\section*{I.~~~~Introduction}

In the past two decades, quantum coherence phenomena have been demonstrated to exist and even support the physiological processes in biology \cite{Lambert13,Ai16}, e.g., avian navigation and exciton energy transfer (EET) in natural photosynthesis. The former has been indirectly confirmed by a number of interesting experiments \cite{Maeda08,Zapka09,Kominis09}. And the entanglement between the pair of natural qubits, radicals, was shown to last over milliseconds at the ambient condition, which is significantly longer than those artificial quantum systems at a sufficiently-low temperature and vacuum \cite{Cai10,Cai11,Vedral11}. In order to unravel the underlying physical mechanism,
Ritz et al. \cite{Ritz00} proposed the radical-pair hypothesis to describe how birds utilize the weak geomagnetic field for navigation. In the radical-pair hypothesis, the inter-conversion between the spin-singlet and triplet states induced by the geomagnetic field and the local field posed by the nuclear spins results in distinguishable products of chemical reaction. This kind of magnetic-field sensitive chemical reactions can be well described by the generalized Holstein model with spin degrees of freedom taken into consideration \cite{Yang12}. Since the detection sensitivity subtly depends on the interplay of the nuclear spins and geomagnetic field, it was suggested that when there is quantum phase transition in the nuclear spins \cite{Quan06,Ai08,Ai09}, the sensitivity can be significantly increased by quantum criticality \cite{Cai12}.

On the other hand, photosynthesis is a complex biochemical reaction process. It is well known that the process of photosynthesis mainly includes four aspects, namely, primary reactions, electron transfer, photophosphorylation and carbon dioxide fixation \cite{Fleming94}. In the process of primary reaction, one of the peripheral light-harvesting antennas absorbs a photon of sunlight, and then the pigments transfer the captured energy to the reaction center.
In the reaction center, electrons are transferred and a potential difference is generated to drive the subsequent biochemical reactions.
The processes of energy and electron transfer are ultrafast, both of them occurring around $10^{-12}$ seconds. The excitation energy is delivered to the reaction center on a time scale of 30 picoseconds, and subsequent electron transfer in about 3 picoseconds. Due to nearly $100\%$ quantum efficiency, the energy transfer mechanism has aroused widespread interests. One of the current research hotspots is the study of the physical mechanism of EET in photosynthesis.
In the late 1980s, the molecular structure of the protein complex of the purple bacteriological reaction center was obtained \cite{Deisenhofer89}. Subsequently, some other molecular structures of reaction centers and light-harvesting antenna protein complexes were generally determined \cite{Deisenhofer91,Mostame12,Potocnik17,Cheng06,Tao18}. The increased understanding of the molecular structures of photosynthesis is helpful to the development of artificial light-harvesting devices \cite{Xu18}.


Up to now, the study on EET has been made much progress \cite{Mostame12,Potocnik17,Tao18,Gorman18,Chin18,Engel07,Cheng06,Xu18,Caruso09,Sarovar11,Fassioli10,Eisfeld12,Leon-Montiel13,Mohseni08,Rebentrost08,Rebentrost09,Cao09,Sachdev99,Tao16,Zhang98,Olaya-Castro08,Hu97,Tan14,Yang10,Liang10}. A series of interesting experiments point out that quantum coherence may play an important role in the process of EET in photosynthesis. For example, it has been demonstrated that exciton delocalization optimizes the energy transfer from B800 ring to B850 ring in LH2 \cite{Cheng06}. In addition, it seems that there is some relationship between quantum coherence and the high efficiency of EET \cite{Engel07}.
Moreover, Caruso et al. \cite{Caruso09} studied the relation between the degree of entanglement and EET efficiency in photosynthesis. They believed that the EET in photosynthesis is realized by the interplay of coherent and incoherent processes.
Sarovar et al. \cite{Sarovar11} found that the coherence between pigments almost exists in the whole process of EET. Fassioli and Olaya-Castro \cite{Fassioli10} showed that there is an inverse
relationship between the quantum efficiency and the average entanglement
between distant donor sites in FMO complex.


As photosynthetic complexes inevitably interact with the surrounding environment, some scientists have also begun to study the quantum noise effects on EET efficiency.
After studying the impact of environment on EET, Leon-Montiel and Torres \cite{Leon-Montiel13} found that even in purely-classical systems, the efficiency of EET can be improved by adjusting the external environment. Mohseni et al. \cite{Mohseni08,Rebentrost08,Rebentrost09} proposed the concept of environmental-noise-assisted EET in the study of FMO photosynthetic complexes.  Cao and Silbey \cite{Cao09} pointed out that the optimal exciton trapping scheme can be achieved by choosing the appropriate decoherence rate and exciton trapping rate. Plenio and his collaborators \cite{Caruso09} revealed that there is a certain correlation among the EET efficiency, the energy detuning and dephasing. Although the above research results were obtained using numerical methods, these results have revealed that adjusting the environmental noise can improve the EET efficiency.

In the above continuous study of EET in photosynthesis, many theoretical methods have been proposed. They can be divided into coherent and incoherent theories. Traditionally, the incoherent theories, e.g., F\"{o}rster theory and MRT, were employed to calculate the energy transfer efficiency. In 2007, Fleming's group \cite{Engel07} first observed quantum coherence in FMO complexes at low temperatures ($T\!\!=\!\!77$~K) for 660~fs. This discovery made people more aware that the energy transfer among pigments does not fully follow F\"{o}rster theory. For the coherent theories, there are hierarchical equation of motion (HEOM), quantum path integrals, coherent modified Redfield theory (CMRT), small-polaron quantum master equation, and general Bloch-Redfield theory. This paper mainly introduces these incoherent and coherent theories, and analyses the applicable conditions, advantages and disadvantages of each theory.

\section*{II.~~~~Incoherent theories for photosynthesis}

In this section, we will mainly introduce the incoherent theories, which are efficient for calculating the efficiency of EET in photosynthesis \cite{Scholes01,Yang02,Jang02-1,Ishizaki09}.

\subsection{F\"{o}rster theory}

Initially, people used F\"{o}rster theory to describe energy transfer in photosynthesis. The condition of F\"{o}rster theory is that the distance between molecules is long enough in the process of energy transfer.
For a donor (D) and an acceptor (A), the rate of EET between them can be described by the F\"{o}rster theory as \cite{May04,Mukame1999}
\begin{equation}
k_{\rm{D}\rightarrow {\rm A}}^{\rm{F}}=\frac{J_{\rm{DA}}^{2}}{2\pi\hbar^{2}}\int_{-\infty}^{\infty}{\rm d}\omega E_{\rm{D}}\left(\omega\right)I_{{\rm A}}\left(\omega\right),\label{eq:FosterRate}
\end{equation}
where $\omega$ is the frequency and $\hbar$ is the Planck constant, $J_{\rm{DA}}$ represents the Coulomb coupling between D and A, which is inversely proportional to the cubic power of the distance between D and A. $E_{\rm{D}}\left(\omega\right)$ and $I_{{\rm A}}\left(\omega\right)$ represent the fluorescence and the absorption spectrum respectively, which can be directly measured by the experiment and can also be calculated from the theory as
\begin{eqnarray}
E_{\rm{D}}\left(\omega\right)&=&\int_{-\infty}^{\infty}{\rm d}t {\rm e}^{-{\rm i}\omega t}\rm{Tr}_{\rm{B}}\left( \rho_{\rm{D}}^{e}{\rm e}^{-{\rm i}H_{\rm{D}}^{e}t/\hbar} {\rm e}^{{\rm i}(\varepsilon_{\rm{D}}+H_{\rm{D}}^{\rm{g}})t/\hbar}\right),        \\
I_{{\rm A}}\left(\omega\right)&=&\int_{-\infty}^{\infty}{\rm d}t {\rm e}^{{\rm i}\omega t}\textrm{Tr}_{\rm{B}}\left( \rho_{{\rm A}}^{\rm{e}}{\rm e}^{{\rm i}H_{{\rm A}}^{\rm{e}}t/\hbar} {\rm e}^{-{\rm i}(\varepsilon_{{\rm A}}+H_{{\rm A}}^{\rm{g}})t/\hbar}\right).
\end{eqnarray}
Here $\varepsilon_{\rm{D}}$ and $\varepsilon_{{\rm A}}$ are excited-state energies of D and A, respectively.
\begin{eqnarray}
H_{\rm{D}}^{\rm{g}}=\sum_j\omega_ja_j^\dagger a_j,        \\
H_{{\rm A}}^{\rm{g}}=\sum_j\omega_jb_j^\dagger b_j,
\end{eqnarray}
are the Hamiltonians corresponding to the baths of D and A when they are in the ground states, respectively. $a_j$ ($b_j$) is the annihilation operator of donor's (acceptor's) $j$th harmonic oscillator with frequency $\omega_j$.
\begin{eqnarray}
H_{\rm{D}}^{\rm{e}}=\sum_j\omega_ja_j^\dagger a_j+\sum_jg_j^{\rm{D}}(a_j^\dagger+a_j),        \\
H_{{\rm A}}^{\rm{e}}=\sum_j\omega_jb_j^\dagger b_j+\sum_jg_j^{{\rm A}}(b_j^\dagger+b_j),
\end{eqnarray}
are the Hamiltonians of D's and A's bath when they are at the excited state respectively, where $g_j^{\rm{D}}$ ($g_j^{{\rm A}}$) is the coupling strength with D's (A's) $j$th mode. $\rho_{\rm{D}}^{\rm{e}}={\rm e}^{-\beta H_{\rm{D}}^{\rm{e}}}/\rm{Tr}_{\rm{B}}{\rm e}^{-\beta H_{\rm{D}}^{\rm{e}}}$ ($\rho_{{\rm A}}^{\rm{e}}={\rm e}^{-\beta H_{{\rm A}}^{\rm{e}}}/\rm{Tr}_{\rm{B}}{\rm e}^{-\beta H_{{\rm A}}^{\rm{e}}}$) is the D's (A's) bath's density matrix at the thermal equilibrium, where $\beta=k_{\rm{B}}T$ is the inverse temperature with $k_{\rm{B}}$ and $T$ being the Boltzmann constant and temperature, respectively.

F\"{o}rster theory has been widely used in practical experiments and theories for studying energy transfer at the early stages. However, it is worth noting that the formula is derived under some approximations. Because it is assumed that the intra-system coupling $J_{\rm{DA}}$ is much smaller than the system-bath couplings $g_j^{\rm{D}}$ and $g_j^{{\rm A}}$. Furthermore, for the sake of simplicity,
the intra-system coupling $J_{\rm{DA}}$ is generally calculated by the dipole-dipole approximation. Therefore, it may break down when the distance between D and A is relative small as compared to charge distribution within the D (A).

\subsection{Generalized F\"{o}rster theory}
\label{sec:GFT}

In the preceding subsection, we have introduced F\"{o}rster theory. It was developed for calculating the rate of EET between a donor (D) and an acceptor (A).
However, for a cluster formed by strong couplings of several chromophores, Eq.~(\ref{eq:FosterRate}) can not accurately simulate population dynamics.
This is mainly because the original F\"{o}rster theory treats the internal couplings of the system as a perturbation.
Therefore, when there are some clusters, the original F\"{o}rster theory should be generalized to describe the EET in a multichromophoric system \cite{Jang04,Ai13}.

The spectral overlap in Eq.~(\ref{eq:FosterRate}) is a central feature of the F\"{o}rster theory.
Note that there is no bath's degree of freedom simultaneously coupled to ${\rm D}$ and ${\rm A}$. Under this assumption, the F\"{o}rster theory can be generalized.

For a multichromophoric system, D and A are composed of ${\rm D}_{j}\left(j=1,\dots,N_{\rm D}\right)$ and ${\rm A}_{k}\left(k=1,\dots,N_{\rm A}\right)$, respectively \cite{Jang04}. We suppose that the exciton Hamiltonians is
\begin{eqnarray}
H_{{\rm D},0}^{\rm{e}}\!\!&=&\!\!\sum_{j=1}^{N_{{\rm D}}}\epsilon_{{\rm D}_{j}}\vert {\rm D}_{j}\rangle\langle {\rm D}_{j}\vert+\sum_{j\neq j^{'}}\Delta_{jj^{'}}^{{\rm D}}\vert {\rm D}_{j}\rangle\langle {\rm D}_{j^{'}}\vert,\\
H_{{\rm A},0}^{\rm{e}}\!\!&=&\!\!\sum_{k=1}^{N_{{\rm A}}}\epsilon_{{\rm A}_{k}}\vert {\rm A}_{k}\rangle\langle {\rm A}_{k}\vert+\sum_{k\neq k^{'}}\Delta_{kk^{'}}^{{\rm A}}\vert {\rm A}_{k}\rangle\langle {\rm A}_{k^{'}}\vert,
\end{eqnarray}
where $\epsilon_{{\rm D}_{j}}\left(\epsilon_{{\rm A}_{k}}\right)$ is the energy of the excitation state $\vert {\rm D}_{j}\rangle\left(\vert {\rm A}_{k}\rangle\right)$, and $\Delta_{jj^{'}}^{{\rm D}}\left(\Delta_{kk^{'}}^{{\rm A}}\right)$ is the electronic coupling between $\vert {\rm D}_{j}\rangle$ and $\vert {\rm D}_{j^{'}}\rangle$ ($\vert {\rm A}_{k}\rangle$ and $\vert {\rm A}_{k^{'}}\rangle$).
In addition to the above electronic states, all other degrees of freedom will be treated as bath.
If each $\vert {\rm D}_{j}\rangle$  is coupled to a bath operator $\rm{B}_{{\rm D}_{j}}$, the total Hamiltonian for excited ${\rm D}$ can be expressed as $H_{{\rm D}}^{\rm{e}}=H_{{\rm D},0}^{\rm{e}}+\sum_{j=1}^{N_{{\rm D}}}B_{{\rm D}_{j}}\vert {\rm D}_{j}\rangle\langle {\rm D}_{j}\vert+H_{{\rm D}}^{\rm{g}}$. Likewise, the total Hamiltonian for excited ${\rm A}$ is  $H_{{\rm A}}^{\rm{e}}=H_{{\rm A},0}^{\rm{e}}+\sum_{k=1}^{N_{{\rm A}}}B_{{\rm A}_{k}}\vert {\rm A}_{k}\rangle\langle {\rm A}_{k}\vert+H_{{\rm A}}^{\rm{g}}$.
For the assumption that there is no given bath mode couples to both $D$ and $A$, it can be equivalent to $\left[H_{{\rm D}}^{\rm{g}},H_{{\rm A}}^{\rm{g}}\right]=\left[H_{{\rm D}}^{\rm{g}},B_{{\rm A}_{k}}\right]=\left[H_{{\rm A}}^{\rm{g}},B_{{\rm D}_{j}}\right]=0$.
For the multichromophoric system, $N_{\rm D}\times N_{{\rm A}}$ terms contribute to the coupling Hamiltonian, $H_{c}=\sum_{j=1}^{N_{{\rm D}}}\sum_{k=1}^{N_{{\rm A}}}J_{jk}\left(\vert {\rm D}_{j}\rangle\langle {\rm A}_{k}\vert+\vert {\rm A}_{k}\rangle\langle {\rm D}_{j}\vert\right)$, where $J_{jk}$ is the dipole-dipole interaction between ${\rm D}_{j}$ and ${\rm A}_{k}$.

Starting from the Fermi's golden rule \cite{Sumi99,Ma15}, we can obtain the multichromophoric F\"{o}rster rate from the cluster of D to the cluster of A, which is
\begin{equation}
k^{\rm{MC}}\left(t\right)=\sum_{j^{'}j^{''}}\sum_{k^{'}k^{''}}\frac{J_{j'k'}J_{j''k''}}{2\pi \hbar^{2}}\int_{-\infty}^{\infty}{\rm d}\omega E_{{\rm D}}^{j''j'}\left(t,\omega\right)I_{{\rm A}}^{k'k''}\left(\omega\right),
\end{equation}
where
\begin{eqnarray}
&&E_{{\rm D}}^{j''j'}\left(t,\omega\right)\!\!\!\equiv2{\rm Re}[\int_{0}^{t}{\rm d}t' {\rm e}^{-{\rm i}\omega t'}\textrm{Tr}_{{\rm D}}\lbrace {\rm e}^{-{\rm i}H_{{\rm D}}^{\rm{g}}t'/\hbar}  \nonumber   \\
&&\times\langle {\rm D}_{j}\vert {\rm e}^{-{\rm i}H_{{\rm D}}^{e}\left(t-t'\right)/\hbar}\vert {\rm D}_{\hat{\mathbf{e}}}\rangle\langle {\rm D}_{\hat{\mathbf{e}}}\vert\rho_{{\rm D}}^{\rm{g}}{\rm e}^{{\rm i}H_{{\rm D}}^{e}t/\hbar}\vert {\rm D}_{j'}\rangle\rbrace],
\end{eqnarray}
\begin{equation}
I_{{\rm A}}^{k'k''}\left(\omega\right)\equiv\int_{-\infty}^{\infty}{\rm d}t {\rm e}^{{\rm i}\omega t}
\textrm{Tr}_{{\rm A}}\lbrace {\rm e}^{{\rm i}H_{{\rm A}}^{\rm{g}}t/\hbar}\langle {\rm A}_{k'}\vert {\rm e}^{-{\rm i}H_{{\rm A}}^{e}t/\hbar}\vert {\rm A}_{k''}\rangle\rho_{{\rm A}}^{\rm{g}}\rbrace,
\end{equation}
are the matrix elements of $\mathbf{E}_{{\rm D}}\left(t,\omega\right)$ and $\mathbf{I}_{{\rm A}}\left(\omega\right)$.
Here Tr$_{\mu}$ represents the trace over all the bath degrees of freedom for $\mu=\rm{D,~A}$.
For $t<0$, the total system is in the ground state represented by $\rho_{{\rm D}}^{\rm{g}}\rho_{{\rm A}}^{\rm{g}}$.
If the appropriate pulse is selected to excite ${\rm D}$ at $t=0$, the total density operator will be changed to $\rho\left(0\right)=\vert {\rm D}_{\hat{\textbf{e}}}\rangle\langle {\rm D}_{\hat{\textbf{e}}}\vert \rho_{{\rm D}}^{\rm{g}}\rho_{{\rm A}}^{\rm{g}}$, where $\vert {\rm D}_{\hat{\textbf{e}}}\rangle=\mathcal{N}\hat{\textbf{e}}\dot\sum_{j}\mathbf{\mu}_{{\rm D}_{j}}\vert {\rm D}_{j}\rangle$, $\mathcal{N}$ is the normalization constant, $\hat{\mathbf{e}}$ is the polarization vector of the radiation, $\mathbf{\mu}_{{\rm D}_{j}}$ is the transition dipole of $\vert {\rm D}_{j}\rangle$.

It has been demonstrated that the entanglement between donor and bath plays an important role in both the emission spectrum and multichromophoric F\"{o}rster rate \cite{Ma15,Ma15-2,Moix15}.
The coupling between the initial system and the bath will cause the bath to deviate from equilibrium, which will further complicate the emission spectrum. And this deviation will affect the subsequent dynamic evolution, which makes the calculation of multichromophoric F\"{o}rster rate more complicated.
In Ref.~\cite{Ma15}, a full 2nd-order cumulant expansion technique is developed to calculate the spectra and multichromophoric F\"{o}rster rate for both localized and delocalized systems.
The full 2nd-order cumulant expansion technique is reliable for the absorption and emission spectra of localized system. Moreover, for both localized and delocalized systems, the multichromophoric F\"{o}rster rates are close to the exact ones, which can not be obtained by the methods of general second-order theories, such as Redfield theory, MRT, and CMRT.
And for the absorption and emission spectra which are expressed in a full 2nd-order cumulant expansion can reduce to the exact results in the monomer case.

\color{black}{
\subsection{Modified Redfield theory}

\label{sec:MRT}

The F\"{o}rster theory and its generalization are valid in the regime where the intra-system couplings are much weaker than the system-bath couplings. On the contrary,
Redfield theory treats the coupling of exciton and external vibration mode as a perturbation and assumes second-order truncation
when the intra-system couplings prevail.

However, in natural photosynthesis, neither conditions hold as the intra-system couplings are comparable to the system-bath couplings. Therefore, the Redfield theory is modified to describe the EET in natural photosynthesis \cite{Zhang98,Yang02}.
The MRT adopts the same basis, i.e., the exciton basis, as the Redfield theory.
That is to say, when there are closely-connected pigment molecules, the excited state can not be described by the excitation of a single pigment, but its excitation extends to more than one pigment to form an exciton state. The state is a combination of several molecular wave functions, which can be expressed as $\vert\alpha\rangle=\sum_nc^\alpha_n\vert n\rangle$, with $\vert n\rangle$ the excited state of the $n$th pigment.

In this basis, the total Hamiltonian is written as
\begin{equation}
H=H^{{\rm el}}+H^{{\rm ph}}+H^{\rm{el-ph}},
\end{equation}
where
\begin{eqnarray}
H^{{\rm el}}\!\!&=&\!\!\sum_{\alpha}\varepsilon_\alpha\vert\alpha\rangle\langle\alpha\vert,       \\
H^{{\rm ph}}\!\!&=&\!\!\sum_{n,i}\omega_{i}b_{ni}^{\dagger}b_{ni},                        \\
H^{\rm{el-ph}}\!\!&=&\!\!\sum_{\alpha,\beta}\sum_{n,i}\vert\alpha\rangle\langle\beta\vert c_{n}^{\alpha}c_{n}^{\beta*}g_{ni}\omega_{i}\left(b_{ni}+b_{ni}^{\dagger}\right)  \nonumber   \\
&&=\sum_{\alpha,\beta}\vert\alpha\rangle\langle\beta\vert\left(H^{\rm{el-ph}}\right)_{\alpha\beta}.\label{eq:Hel-ph:MRT}
\end{eqnarray}
Here, $\varepsilon_{\alpha}$ is the excitation energy of $\vert\alpha\rangle$, $b_{ni}^{\dagger}$ ($b_{ni}$) is the creation (annihilation) operator of the $i$th phonon mode of $n$th pigment, $\omega_{i}$ is the frequency of the phonon mode, and $g_{ni}\omega_{i}$ is the exciton-phonon coupling constant between the localized electronic excitation on site $n$ and its $i$th phonon mode. Although here we assume individual bath for each pigment, the general correlated baths can also be described by this model. The reorganization energy of site $n$ can be expressed as $\lambda_{n}=\sum_{i}g_{ni}^{2}\omega_{i}$. Finally, the transformation from the site basis to the exciton basis yields the factor $\sum_{n}^{N}c_{n}^{\alpha}c_{n}^{\beta}$, which can be considered as the overlap between exciton wavefunctions
$\vert\alpha\rangle$ and $\vert\beta\rangle$.

The MRT aims at dividing the Hamiltonian into a zeroth-order Hamiltonian and a perturbation part, which correspond to the diagonal and off-diagonal system-bath couplings in the exciton basis, respectively.
The zeroth-order Hamiltonian is
\begin{equation}
H_{0}=H^{{\rm el}}+H^{{\rm ph}}+\sum_{\alpha}\vert\alpha\rangle\langle\alpha\vert\cdot\left(H^{\rm{el-ph}}\right)_{\alpha\alpha},
\end{equation}
and the perturbation Hamiltonian is
\begin{equation}
V=\sum_{\alpha\neq\beta}\vert\alpha\rangle\langle\beta\vert\cdot\left(H^{\rm{el-ph}}\right)_{\alpha\beta}.
\end{equation}

Notice that the diagonal part of $H^{\rm{el-ph}}$, cf. Eq.~(\ref{eq:Hel-ph:MRT}), is included in the zeroth-order Hamiltonian while the off-diagonal part of $H^{\rm{el-ph}}$ is exactly the perturbation Hamiltonian. According to second-order perturbation theory, the transfer rate can be expressed as
\begin{eqnarray}
R_{\alpha\beta}(t)\!\!&=&\!\!2\textrm{Re}\int_0^t {\rm d}\tau F_\beta^*(\tau)A_\alpha(\tau)X_{\alpha\beta}(\tau),
\end{eqnarray}
where
\begin{eqnarray}
A_\alpha(t)\!\!&=&\!\!{\rm e}^{-{\rm i}\varepsilon_\alpha t-g_{\alpha\alpha\alpha\alpha}(t)},\\
F_\alpha(t)\!\!&=&\!\!{\rm e}^{-{\rm i}(\varepsilon_\alpha-2\lambda_{\alpha\alpha\alpha\alpha}) t-g_{\alpha\alpha\alpha\alpha}^*(t)}.
\end{eqnarray}
$A_\alpha(t)$ and $F_\alpha(t)$ are related to the absorption and emission lineshape, respectively.
\begin{eqnarray}
X_{\alpha\beta}(t)\!\!\!&=&\!\!\!{\rm e}^{2[g_{\alpha\alpha\beta\beta}(t)+{\rm i}\lambda_{\alpha\alpha\beta\beta}t]}\nonumber\\
&&\times
[\ddot{g}_{\beta\alpha\alpha\beta}(t)-
(\dot{g}_{\beta\alpha\alpha\alpha}(t)-\dot{g}_{\beta\alpha\beta\beta}(t)-2{\rm i}\lambda_{\beta\alpha\beta\beta})\nonumber\\
&&\times(\dot{g}_{\alpha\beta\alpha\alpha}(t)-\dot{g}_{\alpha\beta\beta\beta}(t)-2{\rm i}\lambda_{\alpha\beta\beta\beta})],
\end{eqnarray}
which is the perturbation-induced dynamical term. And
\begin{eqnarray}
g_{\alpha\beta\gamma\delta}(t)\!\!&=&\!\!\sum_{n,m}c_n^\alpha c_n^{\beta *}c_m^\gamma c_m^{\delta *}g_{nm}(t),\\
{\lambda_{\alpha\beta\gamma\delta}}\!\!&=&\!\!\sum_{n}c_n^\alpha c_n^{\beta*}c_n^\gamma c_n^{\delta*}\lambda_{n}.
\end{eqnarray}
Here the lineshape function is
\begin{eqnarray}
g_{nm}(t)\!\!&=&\!\!\int_0^\infty {\rm d}\omega\frac{J_{nm}(\omega)}{\omega^2}\left\{
\textrm{coth}\left(\frac{\omega}{2k_BT}\right)[1-\cos(\omega t)]\right.\nonumber\\
&&\left.+{\rm i}[\sin(\omega t)-\omega t]\right\},
\end{eqnarray}
and the spectral density is
\begin{eqnarray}
J_{nm}(\omega)=\sum_ig_{ni}g_{mi}\omega_i^2\delta(\omega-\omega_i).
\end{eqnarray}

\begin{figure}[!tpb]   
\begin{center}
\includegraphics[width=4 cm]{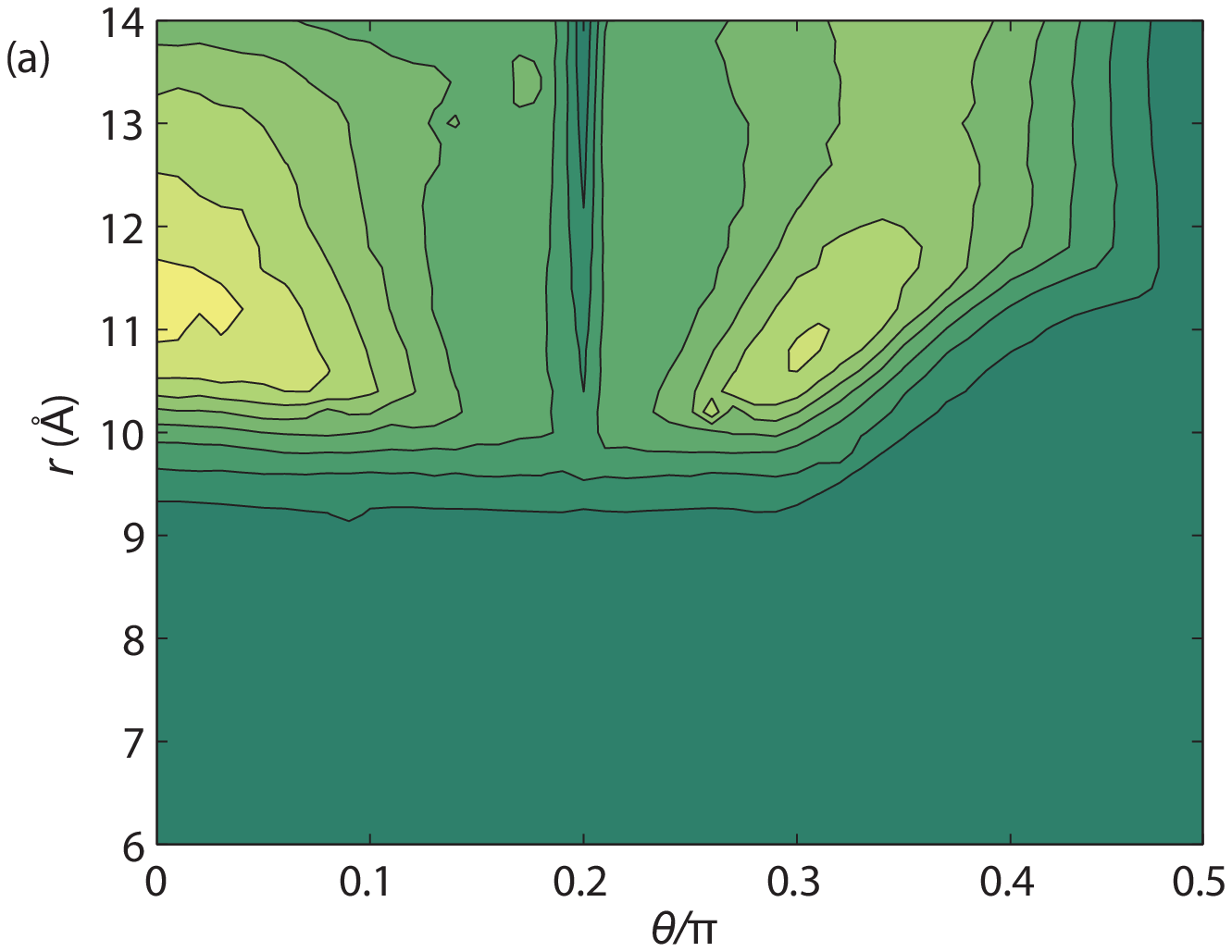}
\includegraphics[width=4 cm]{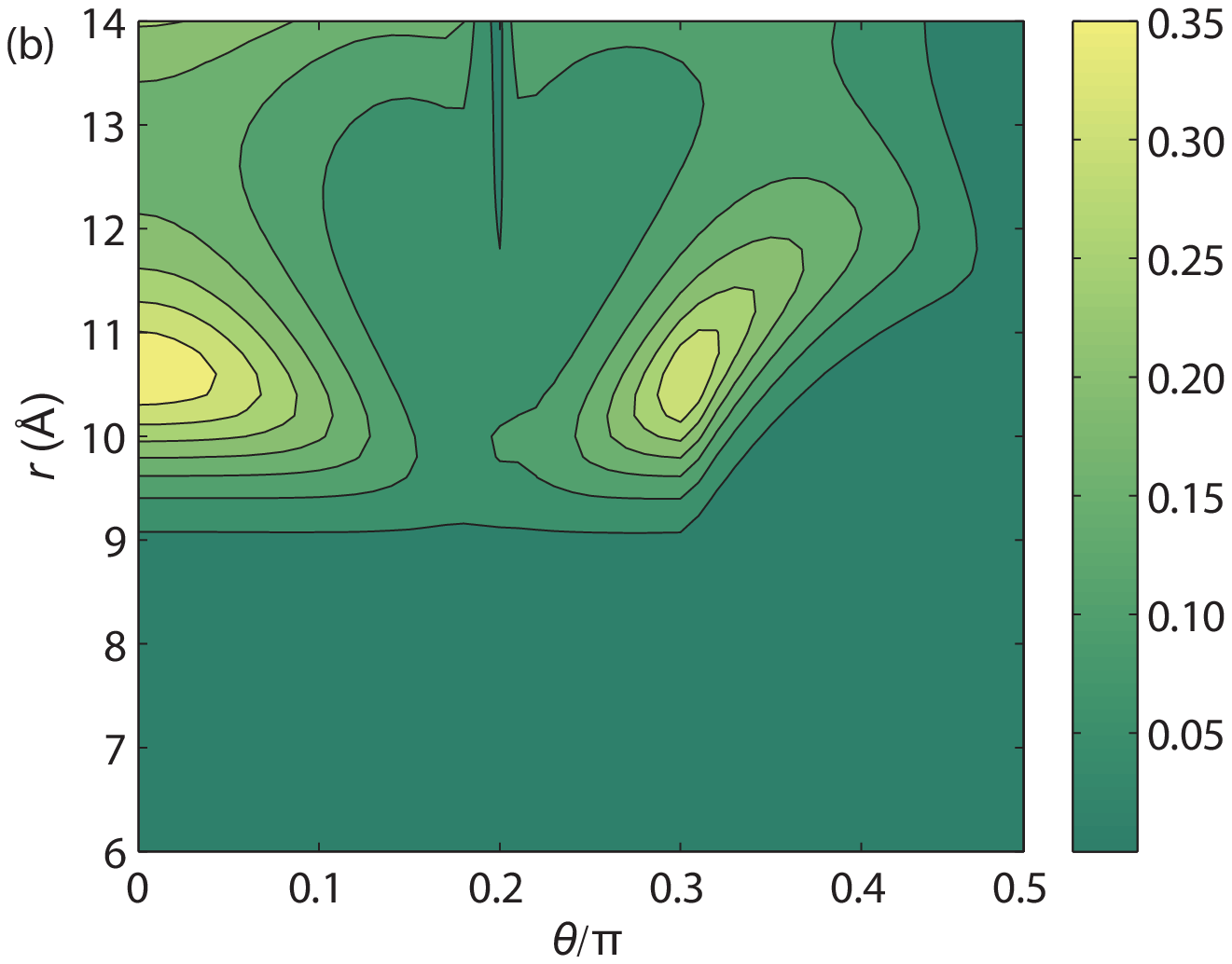}
\caption{
(Color online) Comparison of EET in a tetramer system \cite{Ai13}: The effective transfer rate is obtained by (a) full simulation with the CMRT, (b) cascade exponential decay model with decay rates calculated by the MRT and generalized F\"{o}rster theory.
}\label{figure1}
\end{center}
\end{figure}
}

In Ref.~\cite{Ai13}, in a tetramer system, in addition to the full simulation carried out by the CMRT, the effective transfer rate was also investigated by a cascade exponential decay model with the single decay rate calculated by the MRT within the clusters and generalized F\"{o}rster theory between the clusters, cf. Fig.~\ref{figure1}. It is shown that the simplified calculation captures the effective EET characteristics for a wide parameter range. In this sense, it is reliable to use the incoherent theories to explore the EET efficiency.

It has been shown that the MRT can be successfully used to determine population transfer rates between exciton states beyond the Redfield and F\"{o}rster regimes, because it takes into consideration multi-phonon processes \cite{Yang02}. However, recent studies have demonstrated that the MRT yields unphysical results for small electronic coupling and energy gap, because it underestimates the dynamical localization due to large reorganization energy \cite{Novoderezhkin10,Novoderezhkin13}.

Here, we have summarized the main points of the MRT. Later, by generalizing to describe
the quantum dynamics of off-diagonal terms of density matrix, Cheng et al.
developed the coherent version of the MRT, i.e., the CMRT, to be capable of simulating the quantum dynamics of the full density matrix, which will be introduced in Sec.~\ref{sec:CMRT}.

\section*{III.~~~~Coherent theories for photosynthesis}

Since 2006, there have been a series of interesting experiments \cite{Cheng06,Engel07,Lee07,Collini10,Hildner13}, which implied that coherence might play an important role in the photosynthesis. Since the incoherent theories are only capable of calculating the energy transfer efficiency, theories other than the incoherent ones are needed to consider the effect of coherence on the quantum dynamics. In this section, we summarize the main theories, which can calculate the quantum dynamics of the full density matrix, including both the populations and the off-diagonal terms.

\subsection*{A.~~~~Redfield theory}

For open quantum systems, one of the main concerns is to obtain the equation of the system's reduced density matrix by eliminating the degrees of freedom of the environment \cite{Breuer04,Caldeira83,Grabert88}. Generally, the quantum dynamics of the reduced density matrix can be obtained from the Liouville equation by the perturbation expansion, e.g., Redfield theory \cite{May04}.

For an open quantum system, the total Hamiltonian $\mathbf{H}_{\rm{t}}$ can be expressed as
\begin{equation}
\mathbf{H}_{\rm{t}}=\mathbf{H}_{\rm{S}}+\mathbf{H}_{\rm{B}}+\mathbf{H}_{\rm{SB}},
\end{equation}
where $\mathbf{H}_{\rm{S}}=\sum_{\mu}E_{\mu}\vert\mu\rangle\langle\mu\vert$ represents the Hamiltonian of system with eigen-state $\vert\mu\rangle$ and eigen-energy $E_{\mu}$, $\mathbf{H}_{\rm{B}}=\sum_{n,k}\omega_ka_{nk}^\dagger a_{nk}$ corresponds to the environment, and $\mathbf{H}_{\rm{SB}}=\sum_{n,k}g_k\vert n\rangle\langle n\vert(a_{nk}^\dagger+a_{nk})$ describes the interaction between the system and environment. For the reduced density matrix, $\rho_{\rm{S}}\left(t\right)$ can be expressed as $\rho_{\rm{S}}\left(t\right)=\textrm{Tr}_{\rm{B}}\left[\rho_{\rm{t}}\left(t\right)\right]$, where $\rho_{\rm{t}}\left(t\right)$ is the density matrix of the total system.

By treating $\mathbf{H}_{\rm{S}}$ exactly and $\mathbf{H}_{\rm{SB}}$ to the second order, we can derive a time-local master equation for the open system's density matrix $\rho_{\rm{S}}\left(t\right)$ as
\begin{equation}
\frac{{\rm d}}{{\rm d} t}\rho_{\rm{S}}^{\mu\nu}(t)=-{\rm i}\omega_{\mu\nu}\rho_{\rm{S}}^{\mu\nu}+\sum_{\mu'\nu'}R_{\mu\nu,\mu'\nu'}\rho_{\rm{S}}^{\mu'\nu'},
\label{eq:Redfield}
\end{equation}
where $\omega_{\mu\nu}=(E_\mu-E_\nu)/\hbar$ is the transition frequency from the eigen-state $\vert\mu\rangle$ to $\vert\nu\rangle$, $R_{\mu\nu,\mu'\nu'}$ is the transfer rate from $\rho_{S}^{\mu'\nu'}$ to $\rho_{\rm{S}}^{\mu\nu}$.
It is also called the Redfield tensor,
\begin{eqnarray}
R_{\mu\nu,\mu'\nu'}&=&\Gamma_{\nu'\nu,\mu\mu'}+\Gamma_{\mu'\mu,\nu\nu'}^\ast
-\delta_{\nu'\nu}\sum_{\kappa}\Gamma_{\mu\kappa,\kappa\mu'}\nonumber\\
&&-\delta_{\mu'\mu}\sum_{\kappa}\Gamma_{\nu\kappa,\kappa\nu'}^\ast,
\end{eqnarray}
where the damping matrix is
\begin{equation}
\Gamma_{\mu\nu,\mu'\nu'}=\frac{1}{\hbar^2}\sum_{m,n}\langle\mu\vert V_m\vert\nu\rangle\langle\mu'\vert V_n\vert\nu'\rangle C_{mn}(\omega_{\nu'\mu'}),
\end{equation}
$C_{mn}(\omega_{\nu'\mu'})$ is the Fourier transform of bath correlation function
\begin{equation}
C_{mn}(t)=\frac{\hbar}{\pi}\int_{-\infty}^{\infty}{\rm d}\omega J_{mn}(\omega)(n_{\rm{BE}}+1){\rm e}^{-{\rm i}\omega t}
\end{equation}
with $n_{\rm{BE}}(\omega)=[\exp(\beta\hbar\omega)-1]^{-1}$  the Bose-Einstein distribution function
and $J_{mn}(\omega)$ the spectral density. Here $\beta$ is inversely proportional to temperature.

Redfield theory is traditionally used in calculating the full quantum dynamics of the system's density matrix. Since it treats the system-bath couplings perturbatively,
it is only valid when the intra-system couplings are much stronger than the system-bath couplings. Because the photosynthesis is in the intermediate regime, other approaches should be explored to describe the quantum dynamics of EET in photosynthesis \cite{Ishizaki09,Ishizaki09one}.

\subsection*{B.~~~~Hierarchical Equation of Motion}

The HEOM formalism has become an important tool for studying dynamics of open quantum systems \cite{Tang15,Liu14,Schroder07,Yan13}. And it can be used to study the dynamics in different fields, such as chemistry \cite{Xu13}, biology \cite{Chen11} and physics \cite{Zheng08one,Zheng08two,Zheng09one,Zheng09two}.
The establishment of the HEOM starts from the influence of the functional path
integral \cite{Feynman1963,Weiss09,Kleinert09,Y. Tanimura89,Y. Tanimura90}, and then it is realized by using
the path integration algorithm \cite{Xu05,Xu07,Jin08} and decomposing the environmental correlation functions appropriately \cite{Zheng09two,Xu07,Jin08,Jin07,Hu10,Hu11}.
The HEOM couples the system's reduced density operator and a series
of auxiliary density operators \cite{Xu05,Xu07,Jin08}.

In the following, we describe in detail the theoretical approach of HEOM for
photosynthetic EET \cite{Ishizaki09one,Ishizaki09two}.
For a photosynthetic complex containing $N$ pigments, we can study its EET dynamics by using the Frenkel exciton Hamiltonian \cite{Renger01,Yang02,Cheng09}
\begin{eqnarray}\label{HEOMeq01}
H_{\rm{tot}}=H^{{\rm el}}+H^{{\rm ph}}+H^{\rm{el-ph}},
\end{eqnarray}
where
\begin{eqnarray}
H^{{\rm el}}&=&\sum_{j=1}^{N}\vert j\rangle\varepsilon_{j}\langle j\vert+\sum_{j\neq k}\vert j\rangle J_{jk}\langle k\vert,  \label{HEOMeq02}       \\
H^{{\rm ph}}&=&\sum_{j=1}^{N}H_{j}^{{\rm ph}},         \label{HEOMeq03}      \\
H^{\rm{el-ph}}&=&\sum_{j=1}^{N}H_{j}^{\rm{el-ph}}=\sum_{j=1}^{N}V_{j}u_{j}.     \label{HEOMeq04}
\end{eqnarray}
In the above, $\vert j\rangle$ represents the state where only the $j$th
pigment is in its excited state and
all others are in their ground state, i.e.,
$\vert j\rangle=\vert\varphi_{1g}\rangle\vert\varphi_{2g}\rangle\cdots\vert\varphi_{je}\rangle\cdots\vert\varphi_{Ng}\rangle$.
$\varepsilon_{j}=\varepsilon_{j}^{0}+\lambda_{j}$ is the site energy of the $j$th pigment, where
$\varepsilon_{j}^{0}$ is the excited-state energy of the $j$th site in the
absence of phonons and $\lambda_{j}$ is the reorganization energy of the $j$th site.
$J_{jk}$ is the electronic coupling
between the $j$th and $k$th pigments. In Eq.~(\ref{HEOMeq03}),
$H_{j}^{{\rm ph}}=\sum_{\xi}\omega_{\xi}\left(p_{\xi}^{2}+q_{\xi}^{2}\right)/2$
is the Hamiltonian of the environmental phonons associated with the $j$th pigment,
with $q_{\xi}$ and $p_{\xi}$ dimensionless coordinate and conjugate momentum of the $\xi$th mode. Eq.~(\ref{HEOMeq04}) is the coupling
Hamiltonian between the $j$th site and the phonon modes, with
$V_{j}=\vert j\rangle\langle j\vert$ and $u_{j}=-\sum_{\xi}c_{j\xi}q_{\xi}$,
where $c_{j\xi}$ being the coupling constant between the $j$th pigment and
$\xi$th phonon mode.


The EET dynamics is given by the reduced density matrix
\begin{equation}
\rho^{{\rm el}}\left(t\right)={\rm Tr_{ph}}\{\rho^{{\rm tot}}\left(t\right)\},
\end{equation}
where $\rho^{{\rm tot}}(t)$ denotes the density matrix for the total
system.
In the interaction picture with respect to $H_{0}=H^{{\rm el}}+H^{{\rm ph}}$, the time evolution of the total density matrix $\rho^{{\rm tot}}_{{\rm I}}(t)$ is governed by the Liouville equation,
\begin{equation}
\partial_{t}\rho^{{\rm tot}}_{{\rm I}}(t)=-{\rm i}\mathcal{L}_{{\rm el-ph}}(t)\rho^{{\rm tot}}_{{\rm I}}(t), \label{eq:2}
\end{equation}
where $\mathcal{L}_{{\rm el-ph}}(t)$ is the Liouville operator corresponding to the $H^{{\rm el-ph}}$, which satisfies
$\mathcal{L}_{{\rm el-ph}}(t)\rho^{{\rm tot}}_{{\rm I}}=\left[\mathbf{H}^{{\rm el-ph}}(t),\rho^{{\rm tot}}_{{\rm I}}\right]$. And $H^{\rm{el-ph}}(t)=V_{j}(t)u_{j}(t)$, with $V_{j}(t)=\exp({\rm i}H^{{\rm el}}t)V_{j}\exp(-{\rm i}H^{{\rm el}}t)$, $u_{j}(t)=\exp({\rm i}H^{{\rm ph}}t)u_{j} \exp(-{\rm i}H^{{\rm ph}}t)$.

We suppose that the system and its environment is uncorrelated at the initial time \cite{May04},
i.e.,
$\rho^{{\rm tot}}_{\rm{I}}\left(0\right)=\rho^{{\rm el}}_{\rm{I}}\left(0\right)\rho_{\rm{eq}}^{{\rm ph}}$
with $\rho_{\rm{eq}}^{{\rm ph}}=\exp(-\beta H^{{\rm ph}})/\rm{Tr}_{{\rm ph}}\exp(-\beta H^{{\rm ph}})$ and $\beta=1/k_{\rm B}T$.

By calculating the time integral of Eq.~(\ref{eq:2}), the formal solution of reduced density matrix can be obtained as
\begin{equation}
\rho^{{\rm el}}_{\rm{I}}(t)=\rm{Tr}_{{\rm ph}}\left\lbrace\Gamma_{+}\exp\left[-i\int_{0}^{t}{\rm d}\tau \mathcal{L}_{\rm{el-ph}}(\tau)\right]\rho^{{\rm ph}}_{\rm{eq}}\right\rbrace\rho^{{\rm el}}_{\rm{I}}(0), \label{eq:3}
\end{equation}
where $\Gamma_{+}$ is the forward time-ordering operator. The average over the linear bath operator $u_{j}(t)$ results in ${\rm Tr}_{{\rm ph}}\lbrace u_{j}(t)\rho^{{\rm ph}}_{{\rm eq}}\rbrace=0$ and ${\rm Tr}_{{\rm ph}}\lbrace u_{j}(t)u_{j}(0)\rho^{{\rm ph}}_{{\rm eq}}\rbrace\equiv C(t)\equiv C_{\rm{real}}(t)+{\rm i}C_{\rm{imag}}(t)$, where $C_{\rm{real}}(t)$ and $C_{\rm{imag}}(t)$ are the real and imaginary parts of the correlation function of bath, respectively. Consequently, Eq.~(\ref{eq:3}) is simplified to be
\begin{equation}
\rho^{{\rm el}}_{\rm{I}}(t)=U(t)\rho^{{\rm el}}_{\rm{I}}(0), \label{eq:4}
\end{equation}
with the time propagator of the reduced density matrix
\begin{equation}
U(t)=\Gamma_{+}\exp\left[-\int_{0}^{t} W(\tau){\rm d}\tau\right],
\end{equation}
where $W(t)=\int_{0}^{t}{\rm d}\tau\mathcal{L}_{\rm{z}}(t)C_{\rm{real}}(t-\tau)\mathcal{L}_{\rm{z}}(\tau)+{\rm i}\mathcal{L}_{\rm{z}}(t)C_{\rm{imag}}(t-\tau)\mathcal{S}_{\rm{z}}(\tau)$,
$\mathcal{L}_{\rm{z}}(t)=\left[V_{j}(t),\dots\right]$,
$\mathcal{S}_{\rm{z}}(t)=\left[V_{j}(t),\dots\right]_{+}$.

The coupling of the $j$th pigment to the environmental
phonons can be specified by the spectral density $J_{j}\left(\omega\right)$. We further assume the spectral density as the overdamped-Brownian oscillator model,
$J_{j}\left(\omega\right)=2\lambda_{j}\gamma_{j}\omega/\left(\omega^{2}+\gamma_{j}^{2}\right)$ with $\gamma_{j}^{-1}$ being the relaxation time and $\lambda_{j}$ the reorganization energy \cite{Wang18}.

Thus, the bath correlation function can be calculated as
\begin{eqnarray}
C(t)=C_{\rm{real}}-{\rm i}C_{\rm{imag}}=\int_{-\infty}^{\infty}\frac{{\rm d}\omega}{\pi}J(\omega)\frac{{\rm e}^{{\rm i}\omega t}}{{\rm e}^{\beta \omega-1}},
\end{eqnarray}
where
\begin{eqnarray}
C_{\rm{real}}(t)\!\!&=&\!\!\int_{-\infty}^{\infty}\frac{{\rm d}\omega}{\pi}J(\omega)\coth\frac{\beta\omega}{2}\cos\omega t, \\
C_{\rm{imag}}(t)\!\!&=&\!\!-\int_{-\infty}^{\infty}\frac{{\rm d}\omega}{\pi}J(\omega)\sin\omega t.
\end{eqnarray}

By unitary transformation, we can obtain the reduced density matrix  $\rho^{{\rm el}}(t)$ in the Schr\"{o}dinger picture as $\rho^{{\rm el}}(t)=U_{{\rm el}}(t)\rho^{{\rm el}}_{\rm{I}}(t)U_{{\rm el}}^\dagger(t)=\exp(-{\rm i}H^{{\rm el}}t)\rho^{{\rm el}}_{\rm{I}}(t)\exp({\rm i}H^{{\rm el}}t)$. And its time derivative can be expressed as
\begin{equation}
\partial_{t}\rho^{{\rm el}}(t)=-{\rm i}\mathcal{L}_{\rm{S}}\rho^{{\rm el}}(t)-{\rm i}\mathcal{L}_{\rm{z}}\sigma_{1}(t),
\end{equation}
where $\mathcal{L}_{\rm{S}}=\left[H^{{\rm el}},\dots\right]$ is the commutator of the system Hamiltonian. Here $\sigma_{1}(t)$ is an auxiliary operator, which is
\begin{eqnarray}
\sigma_{1}(t)\!\!\!&=&\!\!\!\Gamma_{+}\lbrace\int_{0}^{t}{\rm d}\tau\left[-{\rm i}C_{\rm{real}}(t-\tau)\mathcal{L}_{\rm{z}}(\tau)+C_{\rm{imag}}(t-\tau)S_{\rm{z}}(\tau)\right]\nonumber\\
&&U_{{\rm el}}(t)U(t)\rbrace\rho^{{\rm el}}(0).
\end{eqnarray}

Under the high-temperature condition, i.e., $\beta\hbar\gamma_{j}<1$,
 ${\rm e}^{\beta\omega}-1\approx\beta\omega$, we can obtain
\begin{equation}
C_{\rm{real}}(t)=\frac{2\lambda_{j}}{\beta}{\rm e}^{-\gamma_{j}t}, C_{\rm{imag}}(t)=-2\lambda_{j}\gamma_{j}{\rm e}^{-\gamma_{j}t}.
\end{equation}
Thus, an arbitrary $n$th-order auxiliary operator can be established as
\begin{eqnarray}
\sigma_{\mathbf{n}}(t)&=&\Gamma_{+}\left\lbrace\int_{0}^{t}{\rm d}\tau_{1}\Theta_{n_{1}}(\tau_{1}){\rm e}^{-\gamma_{j}(t-\tau_{1})}\int_{0}^{t}{\rm d}\tau_{2}\Theta_{n_{2}}(\tau_{2}){\rm e}^{-\gamma_{j}(t-\tau_{2})}\right.\nonumber\\
&&\left.\dots\int_{0}^{t}{\rm d}\tau_{N}\Theta_{n_{N}}(\tau_{N}){\rm e}^{-\gamma_{j}(t-\tau_{N})}U_{{\rm el}}(t)U(t)\right\rbrace\rho^{{\rm el}}(0) \nonumber\\
&&=\Gamma_{+}\lbrace\sum_{j=1}^{N}\left[\int_{0}^{t}{\rm d}\tau\Theta_{n_{j}}(\tau){\rm e}^{-\gamma_{j}(t-\tau)}\right]^{n_{j}}\nonumber\\
&&U_{{\rm el}}(t)U(t)\rbrace\rho^{{\rm el}}(0),\label{eq:sigma n}
\end{eqnarray}
where $\Theta_{n_{j}}(\tau)=-{\rm i}\frac{2\lambda_{j}}{\beta}\mathcal{L}_{\rm{z}}(\tau)-2\lambda_{j}\gamma_{j}S_{\rm{z}}(\tau)$.
Then, we apply the time derivative on both sides of Eq.~(\ref{eq:sigma n}) to derive the equation of motion for the $n$th-order auxiliary operator as
\begin{eqnarray}
\partial_{t}\sigma_{\mathbf{n}}(t)\!\!\!&=&\!\!\!-{\rm i}\mathcal{L}_{\rm{S}}
\sigma_{\mathbf{n}}(t)+\sum_{j=1}^{N}n_{j}(-\gamma_{j})\sigma_{\mathbf{n}}(t)+
\sum_{j=1}^{N}n_{j}\Theta_{n_{j}}(\tau)\sigma_{\mathbf{n}_{j-}}(t)\nonumber\\
&&+\sum_{j=1}^{N}(-{\rm i})\mathcal{L}_{\rm{z}}\sigma_{\mathbf{n}_{j+}}(t), \label{eqHEOM07}
\end{eqnarray}
where $\mathbf{n}=(n_{1},\dots,n_{j},\dots,n_{N})$, $\mathbf{n}_{j\pm}=(n_{1},\dots,n_{j}\pm 1,\dots,n_{N})$ are three sets of nonnegative integers.

Since the hierarchically-coupled Eq.~(\ref{eqHEOM07}) continues to infinity,
we terminate Eq.~(\ref{eqHEOM07}) at a
finite stage as
\begin{equation}
\partial_{t}\sigma_{\mathbf{n}}(t)=-{\rm i}\mathcal{L}_{\rm{S}}\sigma_{\mathbf{n}}(t),
\end{equation}
for the integers $\mathbf{n}=\left(n_{1},n_{2},\ldots,n_{N}\right)$ satisfying
\begin{equation}
\mathcal{N}=\sum_{j=1}^{N}n_{j}\gg\frac{\omega_{e}}{\min\left(\gamma_{1},\gamma_{2},\cdots,\gamma_{N}\right)},
\end{equation}
where $\omega_{e}$ is a characteristic frequency for $\mathcal{L}_{\rm{S}}$. The
number of the operators $\{\sigma\left(\mathbf{n},t\right)\}$ is
$\left(N+\mathcal{N}\right)!/\left(N!\mathcal{N}!\right)$.
Note that Eq.~(\ref{eqHEOM07}) is not suitable for the condition of low-temperature \cite{Ishizaki09two}.

Above all, we only analyze the type of Debye spectral density. It has been proved that HEOM is suitable for any spectral density. Several different methods have been proposed to decompose any spectral density into analytical forms suitable for HEOM processing \cite{Liu14}.
The photosynthetic complex is essentially an open quantum system. Therefore, HEOM can be applied not only to the EET of photosynthesis but also to the related problems in open quantum systems, such as the dynamics of the dissipative electron systems \cite{Jin08}, the non-Markovian entanglement dynamics \cite{Jiang18} and so on.

As demonstrated in Ref.~\cite{Ishizaki09one}, Eq.~(\ref{eqHEOM07}) can describe not only quantum coherent wave-like motion in the Redfield regime, but also incoherent hopping in the F\"{o}rster regime and an
intermediate EET regime in a unified manner. However, although the HEOM numerically exactly produces the quantum dynamics in all three regimes, it takes intolerable computation time, which is exponential in the system's size and the number of exponents in the bath correlation function \cite{Wang18}. Recently, by sophisticated mapping,
we experimentally accelerate the exact simulation of the HEOM in the NMR \cite{Wang18}.

\subsection*{C.~~~~Quantum path integral}

In this subsection, we use a general model to illustrate how path integral can be used in an open quantum system \cite{Thorwart98,Makri95one,Makri95two}. Consider a general quantum system in a bath environment, and assume that the bath is a regular ensemble. The Hamiltonian of the whole system can be expressed as
\begin{equation}
H_{{\rm tot}}=H_{\rm{S}}(t)+H_{\rm{B}}-\sum_{a}Q_{a}F_{a}. \label{QH}
\end{equation}
Here, $H_{\rm{S}}(t)$ is the Hamiltonian of the system. $H_{\rm{B}}$ is the Hamiltonian of the bath. The last term of Eq.~(\ref{QH}) describes the coupling of system and bath, formally decomposed into several dissipative modes, where $Q_{a}$ is the operator of system, and $F_{a}$ is the bath's operator.

The system's reduced density matrix $\rho(t)$ can be defined as
\begin{equation}
\rho(t)={\rm Tr}_{B}\rho_{{\rm tot}}(t)=\mathcal{U}(t,t_{0})\rho(t_{0}).  \label{Qrho}
\end{equation}
Here, $\mathcal{U}(t,t_{0})$ is the propagator.
And Eq.~(\ref{Qrho}) is given at the operator level. However, the expression of path integral must be expanded under certain representation. Consider ${\vert \alpha\rangle}$ to be a set of system's complete bases. Under $\alpha$ representation, assuming that $\bm{\alpha}=(\alpha ,\alpha^{'})$, Eq.~(\ref{Qrho}) can be expressed as ()
\begin{equation}
\rho(\bm{\alpha},t)=\int {\rm d}\bm{\alpha}_{0}\mathcal{U}(\bm{\alpha},t;\bm{\alpha}_{0},t_{0})\rho(\bm{\alpha}_{0},t_{0})
\end{equation}
where $\mathcal{U}(\bm{\alpha},t;\bm{\alpha}_{0},t_{0})$ reads \cite{Feynman1963}
\begin{equation}
\mathcal{U}(\bm{\alpha},t;\bm{\alpha}_{0},t_{0})=\int_{\alpha_{0}[t_{0}]}^{\alpha[t]}{\rm D}\bm{\alpha}{\rm e}^{{\rm i}S[\alpha]}F[\bf{\alpha}]{\rm e}^{-{\rm i}S[\alpha ']}.
\end{equation}
Here, $F$ is the influence functional, which mainly expresses the effect of interaction between the system and bath.
$S[\alpha]$ is the classical action functional along a path $\alpha(\tau)$ between starting point $\alpha(t_{0})=\bm{\alpha}_{0}$ and ending point $\alpha(t)=\bm{\alpha}$. Note that the two endpoints of this path are fixed. If the system-bath interaction is not considered, i.e., $F\!\!=\!\!1$, the dynamics of the system is a completely-coherent process, that is to say, when $F\!\!=\!\!1$, we have $\partial_{t}\mathcal{U}=-i\mathcal{L}\mathcal{U}$, where $\mathcal{L}$ is the Liouville operator of the system, i.e., $\mathcal{L}=[H_{\rm{S}}(t),\dots]$.

Now, the key quantity we consider is the influence functional $F$ induced by the system-bath interaction. Let $\bm{a}=(aa')$ denote a pair of dissipative modes, and introduce the definition \cite{Xu05}
\begin{equation}
\tilde{\mathcal{Q}}(t;{\bm{\alpha}})=\tilde{Q}_{aa'}(t;{\alpha})-\tilde{Q}_{aa'}^{'}(t;{\alpha '}),
\end{equation}
where
\begin{eqnarray}
\tilde{Q}_{\bm{a}}(t;{\alpha})\!\!\!&=&\!\!\!\tilde{Q}_{aa'}=\int_{t_{0}}^{t}{\rm d}\tau C_{\bm{a}}(t-\tau)Q_{a'}[\alpha(\tau)], \\
\tilde{Q}_{\bm{a}}^{'}(t;{\alpha})\!\!\!&=&\!\!\!\int_{t_{0}}^{t}{\rm d}\tau C_{\bm{a}}^{*}(t-\tau)Q_{a'}[\alpha^{'}(\tau)],\\
\mathcal{Q}_{a}[\bm{\alpha}(t)]\!\!\!&=&\!\!\!Q_{a}[\alpha(t)]-Q_{a}[\alpha^{'}(t)].
\end{eqnarray}
Here, for any operator $\hat{O}$, $\mathcal{Q}\hat{O}=[Q_{a},\hat{O}]$. $C_{\bm{a}}(t-\tau)=C_{aa'}(t-\tau)=\langle\hat{F}_{a}(t)\hat{F}_{a'}(\tau)\rangle_{B}$ is the correlation function of the bath.

The final expression of influence functional $F$ reads \cite{Xu05}
\begin{equation}
F[\bm{\alpha}]=\exp\{-\int_{t_{0}}^{t}{\rm d}\tau {R}[\tau ;{\bm{\alpha}}]\},
\end{equation}
where
\begin{equation}
{R}[\tau ;{\bm{\alpha}}]=\sum_{\bm{\alpha}}\mathcal{Q}[\bm{\alpha}(t)]\tilde{\mathcal{Q}}_{a}(t;{\bm{\alpha}}) .
\end{equation}

As a powerful formulism, path integral can be used to calculate quantum dynamics under semi-classical approximation.
When dealing with many-body quantum dynamical problems, path integral does not introduce uncontrolled approximation.
However, in general, the solution of path integral is not easy, because
it should discretize the paths based on time slices in order to generate multi-dimensional integrals.
Therefore, in the long-time limit, the dimension of the integral may be very high.
In addition, numerical techniques for evaluating path integrals in real time and imaginary time are different.

\subsection{Coherent Modified Redfield theory}
\label{sec:CMRT}

We have introduced the MRT in Sec.~\ref{sec:MRT}. In this section, we derive the CMRT by generalizing the MRT. Although the CMRT approach adopts the same basis of the MRT, it can deal with coherent evolution of excitonic systems \cite{Hsien15,Chang15}.

In this section, we adopt the same model as Sec.~\ref{sec:MRT}. In Sec.~\ref{sec:MRT}, we have discussed that the main idea of MRT is to divide the Hamiltonian into a zeroth-order Hamiltonian including the diagonal system-bath couplings in the exciton basis,
\begin{equation}
H_{0}=H^{{\rm el}}+H^{{\rm ph}}+\sum_{\alpha}\vert\alpha\rangle\langle\alpha\vert\cdot\left(H^{\rm{el-ph}}\right)_{\alpha\alpha},
\end{equation}
and the off-diagonal system-bath couplings in the exciton basis as a perturbation,
\begin{equation}
V=\sum_{\alpha\neq\beta}\vert\alpha\rangle\langle\beta\vert\cdot\left(H^{\rm{el-ph}}\right)_{\alpha\beta}.
\end{equation}

Following Born approximation and second-order truncation \cite{Breuer02}, we derive the master equation for coherent EET dynamics,
\begin{eqnarray}
&&\dot{\sigma}_{\alpha\beta}\left(t\right)\!=\!-{\rm i}\left(\varepsilon_{\alpha}-\varepsilon_{\beta}\right)
\sigma_{\alpha\beta}\left(t\right)     \nonumber        \\
&&+\sum_{f}\left[R_{\alpha f}\left(t\right)\sigma_{ff}\left(t\right)-R_{f\alpha}\left(t\right)
\sigma_{\alpha\alpha}\left(t\right)\right]\cdot\delta_{\alpha\beta}     \nonumber    \\
&&-\left[R_{\alpha\beta}^{pd}\left(t\right)+\frac{1}{2}
\sum_{f}\left(R_{f\alpha}\left(t\right)+R_{f\beta}\left(t\right)\right)\right]
\sigma_{\alpha\beta}\left(t\right),\label{eq:sigma:CMRT}
\end{eqnarray}
where
\begin{eqnarray}
R_{\alpha\beta}(t)\!\!\!&=&\!\!\!2{\rm Re}\int_0^t {\rm d}\tau F_\beta^*(\tau)A_\alpha(\tau)X_{\alpha\beta}(\tau),\\
A_\alpha(t)\!\!\!&=&\!\!\!{\rm e}^{-{\rm i}\varepsilon_\alpha t-g_{\alpha\alpha\alpha\alpha}(t)},\\
F_\alpha(t)\!\!\!&=&\!\!\!{\rm e}^{-{\rm i}(\varepsilon_\alpha-2\lambda_{\alpha\alpha\alpha\alpha}) t-g_{\alpha\alpha\alpha\alpha}^*(t)},\\
X_{\alpha\beta}(t)\!\!\!&=&\!\!\!{\rm e}^{2[g_{\alpha\alpha\beta\beta}(t)+{\rm i}\lambda_{\alpha\alpha\beta\beta}t]}\nonumber\\
&&\times
[\ddot{g}_{\beta\alpha\alpha\beta}(t)-
(\dot{g}_{\beta\alpha\alpha\alpha}(t)-\dot{g}_{\beta\alpha\beta\beta}(t)-2{\rm i}\lambda_{\beta\alpha\beta\beta})\nonumber\\
&&\times(\dot{g}_{\alpha\beta\alpha\alpha}(t)-\dot{g}_{\alpha\beta\beta\beta}(t)-2i\lambda_{\alpha\beta\beta\beta})],\\
g_{\alpha\beta\gamma\delta}(t)\!\!\!&=&\!\!\!\sum_{n,m}c_n^\alpha c_n^{\beta *}c_m^\gamma c_m^{\delta *}g_{nm}(t),\\
\lambda_{\alpha\beta\gamma\delta}\!\!\!&=&\!\!\!\sum_{n}c_n^\alpha c_n^{\beta*}c_n^\gamma c_n^{\delta*}\lambda_{n}.
\end{eqnarray}
Here the lineshape function is
\begin{eqnarray}
g_{nm}(t)\!\!\!&=&\!\!\!\int_0^\infty {\rm d}\omega\frac{J_{nm}(\omega)}{\omega^2}\left\{
\rm{coth}\left(\frac{\omega}{2k_BT}\right)[1-\cos(\omega t)]\right.\nonumber\\
&&\left.+{\rm i}[\sin(\omega t)-\omega t]\right\},
\end{eqnarray}
and the spectral density is
\begin{eqnarray}
J_{nm}(\omega)=\sum_ig_{ni}g_{mi}\omega_i^2\delta(\omega-\omega_i).
\end{eqnarray}

\begin{figure}[!h]
\begin{center}
\includegraphics[bb=0 0 220 160,width=8.5 cm,angle=0]{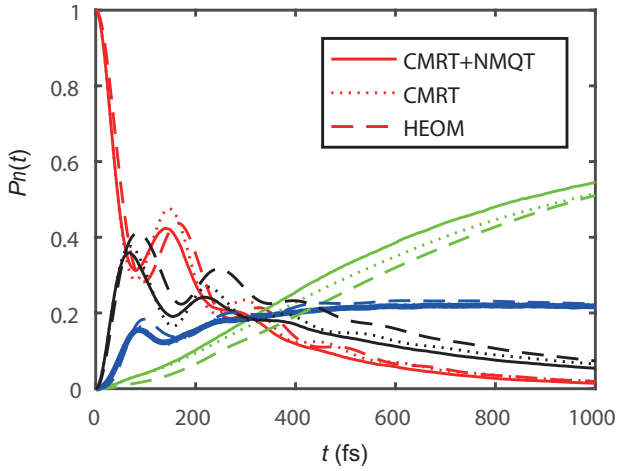}
\caption{
(Color online) $P_n(t)$ vs $t$ of FMO with $P_n(0)=\delta_{n6}$ \cite{Ai14}. The
simulations are performed with $T=77$ K, $\lambda=35~\textrm{cm}^{-1}$, and the effective Hamiltonian from Ref.~\cite{Ishizaki09}. Solid curves are
obtained by the combined Lindblad-form-CMRT-and-NMQJ approach,dotted lines are calculated by the original CMRT, and dashed curves are results by the
HEOM. Red curves are for site 6, green curves for site 3, blue curves for site 4
and black curves for site 5. For the sake of clarity, other sites' population dynamics are small and thus not shown here.
}
\label{figure2}
\end{center}
\end{figure}

In Fig.~\ref{figure2}, we compare the population dynamics by the original CMRT, the HEOM, and the combined approach of Lindblad-form CMRT and NMQT \cite{Ai13,Ai14}. It is shown that the CMRT captures both the coherent oscillations at the short-time regime and the incoherent EET dynamics at the long-time regime. We notice that for sites 6 and 3, the difference between the prediction by the HEOM and the CMRT may be more significant. It was remarked that when the energy gap is small and the exciton-phonon
coupling is strong, the CMRT neglects the dynamical localization. Furthermore, the CMRT-NMQT approach underestimates the coherent evolution, because it might be attributed to the recast the CMRT into the Lindblad form in order to utilize the NMQT approach \cite{Ai13,Ai14}.

By comparison, we can find that the master equation~(\ref{eq:sigma:CMRT}) has the similar form as the traditional Redfield equation~(\ref{eq:Redfield}). Because it is time-local, it is easy to propagate and applicable to large multichromophoric systems. The excitonic Hamiltonian $H^{{\rm el}}$ results in the coherent dynamics as described by the first term in Eq.~(\ref{eq:sigma:CMRT}). The off-diagonal exciton-phonon couplings in the exciton basis $V$ lead to the incoherent population transfer, cf. the second term therein. The last term stands for the dephasing processes induced by both the pure-dephasing and the dissipation effects. The CMRT is non-Markovian since we have not invoked the Markovian approximation in deriving Eq.~(\ref{eq:sigma:CMRT}). Moreover, as in Ref.~\cite{Novoderezhkin13}, we would obatain the same absorption spectrum with lifetime broadening effects due to the dephasing terms in Eq.~(\ref{eq:sigma:CMRT}).
The CMRT is applicable to some problems beyond the original MRT, such as time-resolved spectroscopy and control problems \cite{Ai14}.

\subsection{Small-Polaron quantum master equation}

In addition to the methods mentioned above, a new theory has recently been developed, which interpolates between the F\"{o}rster and Redfield theories by using small-polaron transformation \cite{Jang08}.
This theory is based on the second-order perturbative truncation of the renormalized electron-phonon coupling.

We consider a model consisting of a donor $\left({\rm D}\right)$ and an acceptor $\left({\rm A}\right)$ \cite{Jang08}. And we use $\vert \rm{g}\rangle$ to represent the ground state of the D and A, $\vert {\rm D}\rangle$ ($\vert {\rm A}\rangle$) to represent that only D (A) is in the excited state. In this subsection, we only consider the situation when there is only single electronic-excitation.

Initially, we assume that the state of the system is in the ground state $\vert \rm{g}\rangle$ and the bath is in thermal equilibrium.
Then, a laser pulse selectively excites $\vert \rm{g}\rangle$ to $\vert {\rm D}\rangle$. The EET dynamics for $t>0$ can be described by the total Hamiltonian
\begin{equation}
H=H_{\rm{s}}+H_{\rm{sb}}+H_{\rm{b}},
\end{equation}
where $H_{\rm{s}}=H_{\rm{s}}^{\rm{p}}+H_{\rm{s}}^{\rm{c}}$ represents the system Hamiltonian with $H_{\rm{s}}^{\rm{p}}=E_{{\rm D}}\vert {\rm D}\rangle\langle {\rm D}\vert+E_{{\rm A}}\vert {\rm A}\rangle\langle {\rm A}\vert$ and $H_{\rm{s}}^{\rm{c}}=J\left(\vert {\rm D}\rangle\langle {\rm A}\vert+\vert {\rm A}\rangle\langle {\rm D}\vert\right)$, $J$ is the coupling between $\vert {\rm D}\rangle$ and $\vert {\rm A}\rangle$, $E_{{\rm D}}$  $\left(E_{{\rm A}}\right)$ is the energy of state $\vert {\rm D}\rangle$  $\left(\vert A\rangle\right)$ relative to ground state $\vert \rm{g}\rangle$,
$H_{\rm{b}}$ is the Hamiltonian of the bath, $H_{\rm{sb}}=B_{{\rm D}}\vert {\rm D}\rangle\langle {\rm D}\vert+B_{{\rm A}}\vert {\rm A}\rangle\langle {\rm A}\vert$ is the system-bath interaction Hamiltonian, with $B_{{\rm D}}$ $\left(B_{{\rm A}}\right)$ the bath operator coupled to $\vert {\rm D}\rangle$  $\left(\vert {\rm A}\rangle\right)$. Here, we assume a spin-boson-type model,
\begin{eqnarray}
H_{\rm{b}}\!\!&=&\!\!\sum_{n}\hbar\omega_{n}\left(b_{n}^{\dagger}b_{n}+\frac{1}{2}\right),\nonumber\\
B_{{\rm D}}\!\!&=&\!\!\sum_{n}\hbar\omega_{n}g_{nD}\left(b_{n}^{\dagger}+b_{n}\right),\nonumber\\
B_{{\rm A}}\!\!&=&\!\!\sum_{n}\hbar\omega_{n}g_{nA}\left(b_{n}^{\dagger}+b_{n}\right),
\end{eqnarray}
where $b_{n}^{\dagger}$ $\left(b_{n}\right)$ is the creation (annihilation) operator of the $n$th mode, and its corresponding frequency $\omega_{n}$. According to the initial condition,  $\rho\left(0\right)=\sigma\left(0\right){\rm e}^{-\beta H_{\rm{b}}}/Z$, where $\beta=1/k_{\rm{B}}T$, $Z=\textrm{Tr}_{\rm{b}}\lbrace {\rm e}^{-\beta H_{\rm{b}}}\rbrace$, and $\sigma=\vert {\rm D}\rangle\langle {\rm D}\vert$.
The Liouville operator corresponding to the above Hamiltonians can be expressed as $\mathcal{L}$, $\mathcal{L}_{\rm{s}}^{p}$, $\mathcal{L}_{\rm{s}}^{c}$, $\mathcal{L}_{\rm{sb}}$, and $\mathcal{L}_{\rm{b}}$. The evolution of the density operator over time reads
\begin{equation}
\frac{{\rm d}\rho\left(t\right)}{{\rm d}t}\!\!=\!\!-{\rm i}\mathcal{L}\rho\left(t\right)\!\!=\!\!-{\rm i}\left(\mathcal{L}_{\rm{s}}^{\rm{p}}+\mathcal{L}_{\rm{s}}^{\rm{c}}
+\mathcal{L}_{\rm{sb}}+\mathcal{L}_{\rm{b}}\right)\rho\left(t\right).\label{eq:rho:PT}
\end{equation}

As we have discussed in the above subsections, when the coupling to the bath is weak, the Redfield theory can  be employed, while for the strong coupling to the bath, the F\"{o}rster theory is applicable. The method developed below interpolates between these two limits by combining the polaron transformation \cite{Holstein1959,Silbey84,Harris84,Rackovsky1973,Jackson1983} and a quantum master equation formulism up to the second order \cite{Jang02-1}.

Applying the small-polaron transformation generated by $G=\sum_{n}\left(b_{n}^{\dagger}-b_{n}\right)\left(g_{\rm{nD}}\vert {\rm D}\rangle\langle {\rm D}\vert+g_{\rm{nA}}\vert {\rm A}\rangle\langle {\rm A}\vert\right)$ to Eq.~(\ref{eq:rho:PT}), we can obtain the time evolution equation for $\tilde{\rho}\left(t\right)={\rm e}^{G}\rho\left(t\right){\rm e}^{-G}$ as
\begin{equation}
\frac{{\rm d}\tilde{\rho}\left(t\right)}{{\rm d}t}=-{\rm i}\left(\tilde{\mathcal{L}}_{\rm{s}}^{\rm{p}}+\tilde{\mathcal{L}}_{\rm{s}}^{\rm{c}}+\mathcal{L}_{\rm{b}}\right)\tilde{\rho}\left(t\right),
\end{equation}
where $\tilde{\mathcal{L}}_{\rm{s}}^{\rm{p}}$ and $\tilde{\mathcal{L}}_{\rm{s}}^{\rm{c}}$ are the quantum Liouville operators for
\begin{eqnarray}
\tilde{\mathbf{H}}_{\rm{s}}^{p}\!\!&=&\!\!\tilde{E}_{{\rm D}}\vert {\rm D}\rangle\langle {\rm D}\vert+\tilde{E}_{{\rm A}}\vert {\rm A}\rangle\langle {\rm A}\vert,\\
\tilde{\mathbf{H}}_{\rm{s}}^{c}\!\!&=&\!\!J\left(\theta_{{\rm D}}^{\dagger}\theta_{{\rm A}}\vert {\rm D}\rangle\langle {\rm A}\vert+\theta_{{\rm A}}^{\dagger}\theta_{{\rm D}}\vert {\rm A}\rangle\langle {\rm D}\vert\right).
\end{eqnarray}
In the above equations, $\tilde{E}_{{\rm D}}=E_{{\rm D}}-\sum_{n}g_{\rm{nD}}^{2}\hbar\omega_{n}$ and $\tilde{E}_{{\rm A}}=E_{{\rm A}}-\sum_{n}g_{nA}^{2}\hbar\omega_{n}$, $\theta_{{\rm D}}=\exp\{-\sum_{n}g_{\rm{nD}}\left(b_{n}^{\dagger}-b_{n}\right)\}$ and $\theta_{{\rm A}}=\exp\{-\sum_{n}g_{\rm{nA}}\left(b_{n}^{\dagger}-b_{n}\right)\}$, and $\theta_{{\rm D}}^{\dagger}$ and $\theta_{{\rm A}}^{\dagger}$ are their Hermitian conjugates. The initial condition transforms to $\tilde{\rho}\left(0\right)=\sigma\left(0\right)\theta_{{\rm D}}^{\dagger}{\rm e}^{-\beta H_{\rm{b}}}\theta_{{\rm D}}/Z$, which corresponds to the non-equilibrium state of the bath.

In order to derive the quantum master equation, the transformed total Hamiltonian can be divided as $\tilde{H}=\tilde{H}_{0}+\tilde{H}_{1}$. The zeroth-order term is
\begin{equation}
\tilde{H}_{0}=\tilde{H}_{\rm{0,s}}+H_{\rm{b}},
\end{equation}
where
\begin{eqnarray}
\tilde{H}_{\rm{0,s}}\!\!\!&=&\!\!\!\tilde{E}_{{\rm D}}\vert {\rm D}\rangle\langle {\rm D}\vert+\tilde{E}_{{\rm A}}\vert {\rm A}\rangle\langle {\rm A}\vert+J\omega\left(\vert {\rm D}\rangle\langle {\rm A}\vert+\vert {\rm A}\rangle\langle {\rm D}\vert\right),\\
\omega\!\!\!&=&\!\!\!\langle\theta_{{\rm D}}^{\dagger}\theta_{{\rm A}}\rangle=\langle\theta_{{\rm A}}^{\dagger}\theta_{{\rm D}}\rangle={\rm e}^{-\sum_{n}\coth\left(\beta\hbar\omega_{n}/2\right)\delta g_{n}^{2}/2},
\end{eqnarray}
with $\delta g_{\rm{n}}=g_{\rm{nD}}-g_{\rm{nA}}$. The first-order term $\tilde{H}_{1}$ is defined as
\begin{equation}
\tilde{H}_{1}=\tilde{H}_{\rm{s}}^{\rm{c}}-\langle\tilde{H}_{\rm{s}}^{\rm{c}}\rangle=J\left(\tilde{\rm{B}}\vert {\rm D}\rangle\langle {\rm A}\vert+\tilde{\rm{B}}^{\dagger}\vert {\rm A}\rangle\langle {\rm D}\vert\right),
\end{equation}
where $\tilde{\rm{B}}=\theta_{{\rm D}}^{\dagger}\theta_{{\rm A}}-\omega$, $\langle\tilde{\rm{B}}\rangle=\langle\tilde{\rm{B}}^{\dagger}\rangle=0 $. We take $J\tilde{\rm{B}}$ and $J\tilde{\rm{B}}^{\dagger}$ as perturbations, which remain small in both limits of weak and strong system-bath couplings. Thus, the second-order quantum master equation with respect to $\tilde{H}_{1}$ is valid in both limits.

In the interaction picture with respect to $\tilde{H}_{0}$, the time evolution equation of density operator $\tilde{\rho}_{\rm{I}}\left(t\right)=\exp\{{\rm i}\tilde{L}_{0}t\}\tilde{\rho}\left(t\right)$ is
\begin{equation}
\frac{\rm d}{{\rm d}t}\tilde{\rho}_{{\rm I}}\left(t\right)=-{\rm i}\tilde{L}_{\rm{1,I}}\left(t\right)\tilde{\rho}_{{\rm I}}\left(t\right),\label{eq:rhoI:PT}
\end{equation}
where $\tilde{L}_{\rm{1,I}}\left(t\right)$ is the quantum Liouville operator corresponding to
\begin{equation}
\tilde{H}_{\rm{1,I}}(t)=J\left[\tilde{\rm{B}}(t)T(t)+\tilde{B}^{\dagger}(t)T^{\dagger}(t)\right].\label{eq:HI:PT}
\end{equation}
Here, $\tilde{\rm{B}}(t)\!\!\!=\!\!\!\exp\{{\rm i}H_{\rm{b}}t/\hbar\}\tilde{B}\exp\{-{\rm i}H_{\rm{b}}t/\hbar\}$ and $T(t)\!\!\!=\!\!\!\exp\{{\rm i}\tilde{H}_{\rm{0,s}}t/\hbar\}\vert {\rm D}\rangle\langle {\rm A}\vert \exp\{-{\rm i}\tilde{H}_{\rm{0,s}}t/\hbar\}$.

We apply the standard projection-operator technique \cite{Jang08} with $P(\cdot)\equiv\rho_{\rm{b}}\textrm{Tr}_{\rm{b}}\lbrace\cdot\rbrace$ and $Q=1-P$ to Eq.~(\ref{eq:rhoI:PT}), and then make second-order approximations with respect to $\mathcal{L}_{\rm{1,I}}\left(t\right)$ for both the homogeneous and inhomogeneous terms consistently to obtain
\begin{eqnarray}
\frac{\rm d}{{\rm d}t}P\tilde{\mathcal{\rho}}_{\rm{I}}(t)\!\!\!&=&\!\!\!-{\rm i}P\tilde{\mathcal{L}}_{\rm{1,I}}\left(t\right)Q\tilde{\rho}(0)-\int_{0}^{t}{\rm d}\tau P\tilde{\mathcal{L}}_{\rm{1,I}}\left(t\right)\tilde{\mathcal{L}}_{\rm{1,I}}\left(\tau\right) \nonumber   \\
&&   \times\left(Q\tilde{\rho}(t)+P\tilde{\rho}_{\rm{t}}(\tau)\right),
\end{eqnarray}
where $P\tilde{L}_{\rm{1,I}}\left(t\right) P\!\!\!\!=\!\!\!\!0$ and $Q\tilde{\rho}\left(0\right)\!\!\!\!=\!\!\!\!\sigma(0)\times\left(\theta_{{\rm D}}^{\dagger}{\rm e}^{-\beta H_{\rm{b}}}\theta_{{\rm D}}-{\rm e}^{-\beta H_{\rm{b}}}\right)/Z$ has been used. In the above equation, $P\tilde{\rho}_{\rm{I}}\left(\tau\right)$ can be replaced by $P\tilde{\rho}_{\rm{I}}(t)$ without affecting the accuracy up to the second order. By tracing the  bath degrees of freedom to the resulting equation, we obtain the following time-local quantum master equation for $\tilde{\sigma}_{\rm{I}}\left(t\right)=\textrm{Tr}_{\rm{b}}\lbrace\tilde{\rho}_{\rm{I}}\left(t\right)\rbrace$,
\begin{equation}
\frac{\rm d}{{\rm d}t}\tilde{\sigma}_{\rm{I}}\left(t\right)=-\mathcal{R}\tilde{\sigma}_{\rm{I}}\left(t\right)+\mathcal{I}\left(t\right),\label{eq:sigmaI:PT}
\end{equation}
where
\begin{eqnarray}
\mathcal{R}\left(t\right)\!\!\!&=&\!\!\!\int_{0}^{t}{\rm d}\tau {\rm Tr}_{b}\lbrace\tilde{\mathcal{L}}_{\rm{1,I}}\left(t\right)\mathcal{L}_{\rm{1,I}}(\tau)\rho_{\rm{b}}\rbrace,\label{eq:R:PT}\\
\mathcal{I}(t)\!\!\!&=&\!\!\!-{\rm i}{\rm Tr}_{\rm{b}}\lbrace\tilde{\mathcal{L}}_{\rm{1,I}}(t)Q\tilde{\rho}(0)\rbrace \nonumber  \\
&&-\int_{0}^{t}{\rm d}\tau {\rm Tr}_{b}\lbrace\tilde{\mathcal{L}}_{\rm{1,I}}\left(t\right) Q\tilde{\mathcal{L}}_{\rm{1,I}}\left(\tau\right)\tilde{\rho}(0)\rbrace.\label{eq:I:PT}
\end{eqnarray}
Although Eq.~(\ref{eq:sigmaI:PT}) is local in time, the non-Markovian effects can be explained by the time dependence of $\mathcal{R}\left(t\right)$. It's expected to show good performance beyond the typical perturbative regime, as proved for some cases \cite{Palenberg01}. The expressions of the specific time-local quantum master equation obtained by calculation can be referred to in Ref.~\cite{Jang08}.

In Sec.~\ref{sec:GFT}, the effect of system-bath entanglement on the EET has been addressed. Besides, this issue could also be explored by the polaron tranformation \cite{Xu16}. Under certain circumstances, the analytical solutions for the EET efficiency can be explicitly given. It is shown that the results by the polaron transformation are consistent with those by the F\"{o}rster and Redfield theories in the respective limits and interpolate smoothly between them in the intermediate regime.

Beyond the intermediate regime, the small-polaron quantum master equation is accurate in the limits of weak system-bath coupling and weak electronic coupling.
The small-polaron quantum master equation has been applied to many studies, such as exploring the role of the relationship between environmental fluctuation and phonon excitation in EET dynamics \cite{Nazir13}. And it is also used to study the vibrational contributions to EET in photosynthetic complexes of cryptophyte algae \cite{Kolli12}.
Specifically, the small-polaron quantum master equation can be precisely executed in the incoherent region with larger electron-phonon coupling and in the coherent region with faster bath relaxation. In addition, through the special selection of perturbation terms, the weak electron-phonon coupling limit can also be accurately calculated by the small-polaron theory \cite{Chang13}.
However, due to the perturbation, the accuracy of small-polaron quantum master equation declines in the intermediate-coupling regime.

\subsection{Generalized Bloch-Redfield theory}

In photosynthesis, when a light-harvesting pigment absorbs a photon, the photosynthetic system is activated. The excitation energy is then transferred to the reaction center for subsequent charge separation, resulting in energy trapping. Energy decays during transmission and redistributes through interactions with the protein environment.

For the Debye-form spectral density, $J\left(\omega\right)\!\!\!=\!\!\!\left(2\hbar/\pi\right)\lambda\omega D/\left(\omega^{2}+D^{2}\right)$, the spatially-correlated spectral density is $J_{nm}\left(\omega\right)=c_{nm}J\left(\omega\right)$, where $\lambda$ is the reorganization energy, $D$ is the bath relaxation rate, $c_{mn}$ is the spatial correlation coefficient between sites $m$ and $n$. The time correlation functions corresponding to the Debye spectral density can be expanded using exponential functions \cite{Jang02-2}
\begin{equation}
C_{mn}\left(t\right)=c_{mn}\sum_{i=0}^{\infty}\left(f_{i}^{r}+{\rm i}f_{i}^{i}\right){\rm e}^{-\nu_{i}t},
\end{equation}
where $\nu_{i}$ is the relaxation rate of the $i$th bath mode, $f_{i}^{r}$ and $f_{i}^{i}$ are the real and imaginary part of the expansion coefficient, respectively. With the facilitation of auxiliary fields $g_{m;i}\left(t\right)$ at site $m$, the generalized Bloch-Redfield equation for exciton dynamics can be written as \cite{Wu10}
\begin{eqnarray}
\dot{\rho}\left(t\right)\!\!\!&=&\!\!\!-\left(L_{\rm{sys}}+L_{\rm{trap}}\right)\rho\left(t\right)-{\rm i}\sum_{i=0}^{\infty}\sum_{m}\left[Q_{m},g_{m;i}\left(t\right)\right],\\
\dot{g}_{m;i}\left(t\right)\!\!\!&=&\!\!\!-\left(L_{\rm{sys}}+L_{\rm{trap}}+\nu_{i}\right)g_{m;i}\left(t\right)  \nonumber \\
\!\!&&\!\!-{\rm i}f_{i}^{r}\left[\Theta_{m},\rho\left(t\right)\right]+f_{i}^{i}\left[\Theta_{m},\rho\left(t\right)\right]_{+},
\end{eqnarray}
where $\rho$ is the system's reduced density matrix, $Q_{m}=\vert m\rangle\langle m\vert$ is the system operator, $\Theta_{m}=\sum_{n}c_{mn}Q_{n}$, and $\left[A,B\right]_{+}=AB+BA$ is the anti-commutator.
$L_{\rm{sys}}\rho=i\left[\mathbf{H},\rho\right]/\hbar$ and $\mathbf{H}_{nm}=\delta_{n,m}\varepsilon_{n}+\left(1-\delta_{n,m}\right)J_{nm}$, where $\varepsilon_{n}$ is the site energy, the off-diagonal element $J_{nm}$ is the dipole-dipole interaction coupling strength between two distinct sites.
$\left[L_{\rm{trap}}\right]_{nm}\!\!\!=\!\!\!\left(k_{t;m}+k_{t;n}\right)/2$ describes the localization at the charge-separation state, with $k_\textrm{t;m}$ the trapping rate at site $m$. The initial value of $g_{m;i}\left(t\right)$ is zero. The auxiliary field represent the memory effect in the dissipative dynamics caused by the interaction with the bath. For the auxiliary fields, it can be considered as an element of the projection operator $Q\left(1-P\right)$ in the Nakajima-Zwanzig projection operator technique.

As described earlier, the Redfield theory can be applied to the case of weak dissipation regime, but it can not accurately describe exciton dynamics in the case of intermediate and strong dissipation regimes \cite{Rebentrost09,Grover1971}.
The generalized Bloch-Redfield method proposed in this subsection is applicable to a wide range of dissipation cases and can also predict the correct F\"{o}rster-rate limit \cite{Cao97}.

\section*{IV.~~~~Discussion and Conclusion}

Generally, the light-harvesting efficiency in the primary process of photosynthetic organisms can reach up to $95\%$. If we can effectively simulate photosynthesis in nature and learn from it, it will be the most valuable way to solve our current energy problems.

In this paper, we first give a detailed description and summary of the progress in the study of the physical mechanism of EET in photosynthesis. Then, we summarize some theoretical methods for studying incoherent EET in photosynthesis. The F\"{o}rster theory can be used to calculate the rate of EET, but it considers the transfer of electronic excitation energy from a donor to an acceptor. Therefore, the theory should be generalized when applied to a multichromophoric system. In view of this situation, the generalized F\"{o}rster theory has been developed. THe F\"{o}rster theory is applicable to the case where the couplings between pigments are much less than those between the pigments and environment. In this case, the couplings between pigments are regarded as a perturbation. On the contrary, when the couplings between pigments are much greater than those between the pigments and environment, the EET is usually described by the Redfield theory. In this case, the couplings between pigments and environment are regarded as a perturbation. These two theories represent two opposite-limit regions of the coupling strength between pigments and environment. However, because for some intermediate regions, the above two methods are no longer applicable, some new methods and theories are constantly developed, such as the HEOM, which is developed in a non-perturbative manner. The hierarchy equation~(\ref{eqHEOM07}) can describe quantum coherent wave-like motion, and incoherent hopping, and intermediate EET regime in a unified manner. The only disadvantage is that the computational complexity grows exponentially with respect to the system's size and number of exponents in the bath's correlation function. In addition, the quantum master equation by small-polaron transformation has been developed and is used to interpolate between the Redfield and F\"{o}rster limits by employing the small-polaron transformation. It is based on the second-order perturbative truncation with respect to the renormalized electron-phonon coupling. Furthermore, other methods have been developed, such as, quantum path integrals, MRT and its coherent version CMRT, the generalized Bloch-Redfield theory. These methods are also introduced in this paper.

\textbf{Conflict of interest}
The authors declare that they have no conflict of interest.
 
\begin{acknowledgments}
This work was supported by the National Natural Science Foundation of China (11674033, 11474026, and 11505007).
\end{acknowledgments}

\textbf{ Author contributions}
Gui-Lu Long, Qing Ai, and Fu-Guo Deng conceived the review topic. Ming-Jie Tao, Na-Na Zhang, and Qing Ai wrote the manuscript in consultation with all the other authors. Na-Na Zhang arranged all the figures. Peng-Yu Wen and Fu-Guo Deng modified the manuscript. All authors contributed to the final manuscript.

\end{CJK*}

\end{document}